\documentclass[runningheads]{llncs}
\usepackage{graphicx}
\usepackage{booktabs}
\usepackage{multirow}
\usepackage{amssymb}
\usepackage{xcolor}
\begin{document}
\title{Deformed2Self: Self-Supervised Denoising for Dynamic Medical Imaging}
\titlerunning{Deformed2Self}
\author{Junshen Xu\inst{1} \and
Elfar Adalsteinsson\inst{1,2}
}
% index{Xu, Junshen}
% index{Adalsteinsson, Elfar}

%
\authorrunning{J. Xu et al.}
\institute{Department of Electrical Engineering and Computer Science, MIT, \\ Cambridge, MA, USA \\\email{junshen@mit.edu} \and
Institute for Medical Engineering and Science, MIT, Cambridge, MA, USA}

\maketitle              
\setcounter{footnote}{0}

\begin{abstract}
Image denoising is of great importance for medical imaging system, since it can improve image quality for disease diagnosis and downstream image analyses. In a variety of applications, dynamic imaging techniques are utilized to capture the time-varying features of the subject, where multiple images are acquired for the same subject at different time points. Although signal-to-noise ratio of each time frame is usually limited by the short acquisition time, the correlation among different time frames can be exploited to improve denoising results with shared information across time frames. With the success of neural networks in computer vision, supervised deep learning methods show prominent performance in single-image denoising, which rely on large datasets with clean-vs-noisy image pairs. Recently, several self-supervised deep denoising models have been proposed, achieving promising results without needing the pairwise ground truth of clean images. In the field of multi-image denoising, however, very few works have been done on extracting correlated information from multiple slices for denoising using self-supervised deep learning methods. In this work, we propose Deformed2Self, an end-to-end self-supervised deep learning framework for dynamic imaging denoising. It combines single-image and multi-image denoising to improve image quality and use a spatial transformer network to model motion between different slices. Further, it only requires a single noisy image with a few auxiliary observations at different time frames for training and inference. Evaluations on phantom and \textit{in vivo} data with different noise statistics show that our method has comparable performance to other state-of-the-art unsupervised or self-supervised denoising methods and outperforms under high noise levels.

\end{abstract}

\section{Introduction}

Noise is inevitable in medical images. A variety of sources lead to noisy images, such as acquisition with better spatial resolution in MRI~\cite{kale2009trading} and reduction of radiation dose PET~\cite{xu2017200x} and CT~\cite{wolterink2017generative}. Further complicating the task is the complex noise statistics in different medical imaging modalities, which is not limited to additive white Gaussian noise. For example, the noise can be related to pixel intensity like Rician noise in magnitude images of MRI~\cite{you2019denoising} or it can be affected by geometry parameters of the scanner like in CT~\cite{schirrmacher2018temporal}. Thus, a robust method for noise reduction plays an important role in medical image processing and also serves as a core module in many downstream analyses.

In many applications, more than one image is acquired during the scan to capture the dynamic of the subject, e.g., cine images for cardiac MRI~\cite{malayeri2008cardiac}, abdominal dynamic contrast-enhanced (DCE) MRI~\cite{johansson2018abdominal}, and treatment planning 4D thoracic CT~\cite{castillo2009framework}. Image denoising is more necessary for dynamic imaging methods, since they often adopt fast imaging techniques to improve temporal resolution, which may reduce signal-to-noise ratio of each time frame. Dynamic imaging also provides more information for denoising as the images at different time frames have similar content and often follow the same noise model.

In recent years, a number of deep learning based methods have been proposed for image denoising~\cite{zhang2017beyond,chen2018image,lefkimmiatis2018universal,jia2019focnet}. Similar methods are also applied to medical image denoising, e.g., low-dose PET~\cite{xu2020ultra}, CT~\cite{chen2017low} and MRI~\cite{xie2020denoising} denoising. These methods train a convolution neural network (CNN) that maps the noisy image to its clean counterpart. However the supervised training process requires a large-scale dataset with clean and noisy image pairs which can be difficult to acquire, especially in the field of medical imaging. Recently, there are several studies on learning a denoising network with only noisy images. The Noise2Noise~\cite{lehtinen2018noise2noise} method trains the model on pairs of noisy images with the same content but different noises so that the network estimates the expectation of the underlaying clean image. However, in practice, it is difficult to collect a set of different noisy realizations of the same image. The Noise2Void~\cite{krull2019noise2void} and Noise2Self~\cite{batson2019noise2self} methods try to learn denoising network using a dataset of unpaired noisy images and use the blind-spot strategy to avoid learning a identity mapping. Yet, to achieve good performance, these two methods still require the test images to be similar to the training images in terms of image content and noise model. To address this problem, some denoising methods have been developed that only train networks with internal information from a single noisy image and do not rely on large training set. Ulyanov \textit{et al.} introduced deep image prior (DIP) for single-image recovery~\cite{ulyanov2018deep}. Recently, Quan \textit{et al.} proposed Self2Self~\cite{quan2020self2self} method, where it uses dropout~\cite{srivastava2014dropout} to implement the blind-spot strategy.

In terms of denosing methods for dynamic imaging, Benou \textit{et al.} proposed a spatio-temporal denoising network for DCE MRI of brain, where the motion is negligible. Another category of methods first apply conventional registration method to register the images and then perform denoising on the registered images~\cite{schirrmacher2018temporal,lukas2019noise}. However, traditional optimization methods are time consuming~\cite{balakrishnan2019voxelmorph} and registering noisy images directly may reduce the accuracy of registration.

In this work, we propose a deep learning framework for dynamic imaging denoising, named Deformed2Self, where we explore similarity of images in dynamic imaging by deforming images at different time frames to the target frame and utilize the fact that noises of different observations are independent and following similar noise model.
Our method has the following features: 1) The whole pipeline can be trained end-to-end, which is efficient for optimization. 2) Our method is fully self-supervised, i.e., we only need noisy images without ground-truth clean images. 3) The model can be trained on a single image (with a few auxiliary observations) and has no prerequisite on large training dataset, making it suitable for applications with scarce data.

\section{Methods}

\subsection{Problem formulation}

Let $y_0$ be the noisy image we want to denoise, which is generated with some noise models, e.g., for additive noise,
\begin{equation}
    y_0 = x_0 + n_0,
\end{equation}
where $x_0$ denotes the unknown clean image, and $n_0$ denotes the random measurement noise. The goal of single-image denoising is to find a mapping $f$ that can recover $x_0$ from $y_0$, i.e., $\hat{x}_0=f(y_0)\approx x_0$.

In dynamic imaging, multiple images are acquired for the same subject at different time frames. Suppose we have another $N$ frames besides the target frame $y_0$. The noisy observations of these $N$ frames and their unknown clean counterparts are denoted as $\{y_1, ..., y_N\}$ and $\{x_1, ..., x_N\}$ respectively, where $y_k=x_k+n_k$, $k=1,...,N$. For dynamic imaging denoising, information from different time frames are aggregated to estimate the clean image at target frame (frame $0$), $\hat{x}_0=f(y_0, y_1, ..., y_N)$. In many cases, the motion of subject occur during the scan is not negligible. Let $\phi_k$ be the deformation field between the target frame and frame $k$, $x_0=x_k\circ\phi_k$. The noisy observation $y_k$ can be rewritten as
\begin{equation}
    y_k = x_k + n_k = x_0\circ\phi^{-1}_k + n_k, \quad k=0,1,...,N.
    \label{eqn:noise_model}
\end{equation}
where $\phi_0=\mathbf{I}$ is the identity mapping. Eq.~\ref{eqn:noise_model} indicates that the observed noisy images $\{y_k\}_{k=0}^N$ is generated from the target clean image $x_0$ following certain motion models and noise statistics. 
Therefore the auxiliary images from different time frames provide information for estimating the clean image at target frame.

\begin{figure}[t]
\centering
\includegraphics[width=\textwidth]{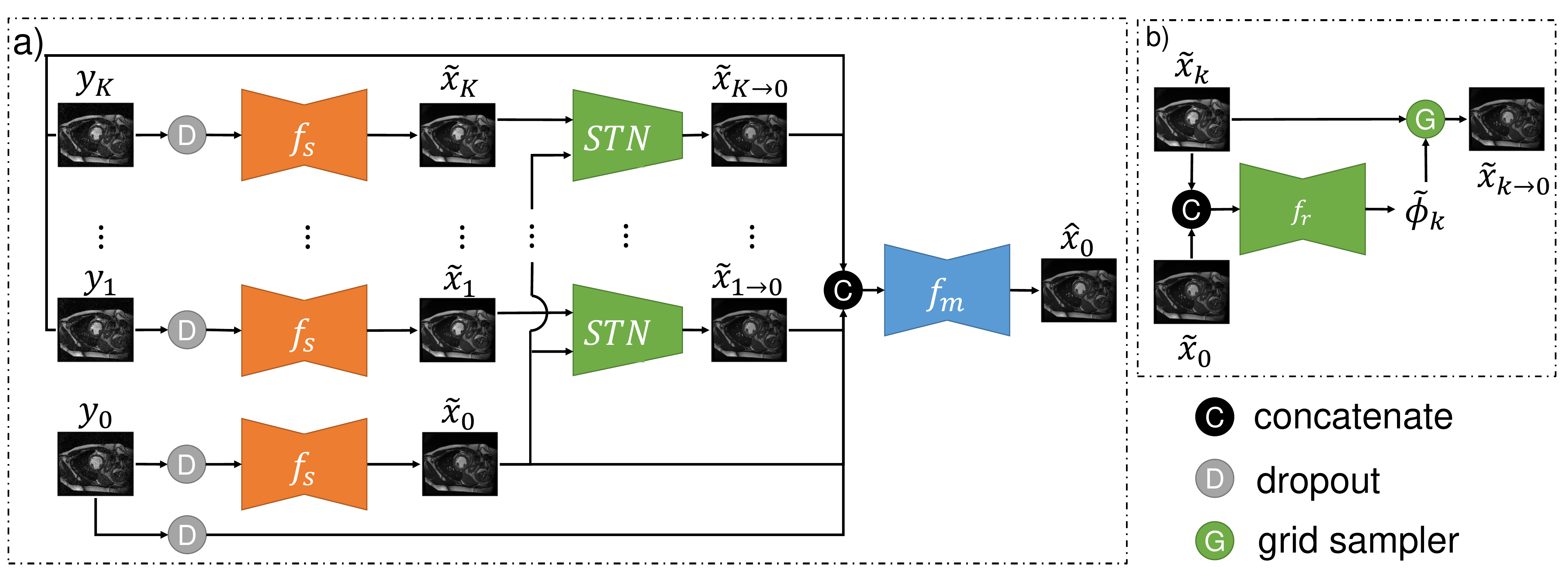}
\caption{Overview of the proposed self-supervised denoising framework. a) The architecture of Deformed2Self, where $f_s$ and $f_m$ are single- and multi-image denoising networks respectively. b) Details of the STN module, where $f_r$ is a registration network.}
\label{fig:method}
\end{figure}

\subsection{The Deformed2Self framework}

Inspired by the data model above, we proposed a self-supervised denoising framework for dynamic imaging named Deformed2Self (Fig.~\ref{fig:method}), which consists of three modules: 1) a single-imaging denoising network to provide coarse estimation of the clean image, 2) a spatial transformer network (STN)~\cite{jaderberg2015spatial} for matching spatial content of images from other time frames $(k=1,...,N)$ to the target frame, and 3) a multi-image denoising network to generate a refined estimation of the clean image at target frame.

{\bf Single-image denoising: }The first step of the proposed framework is single-image denoising, where we use a UNet-based~\cite{ronneberger2015u} network $f_s$ to denoise each frame separately. The benefit of this step is two-fold: 1) It provides a coarse estimation of the clean image utilizing only the information of each frame, which serves as a good initialization for the multi-image denoising to accelerate convergence. 2) It also improves the registration accuracy in the following STN module, since the prediction of deformation field suffers heavily from image noise.

To denoise images with internal information of each frame, we adopt the dropout-based blind-spot strategy~\cite{krull2019noise2void,quan2020self2self}. Given a noisy observation $y_k$, $k\in\{0,...,N\}$, we use a dropout layer to mask out some pixels in $y_k$ before feeding it to the network. The output can be written as $\tilde{x}_{k} = f_s(b_k\odot y_k)$, where $b_k$ is a random binary mask and $\odot$ is the Hadamard product. The network $f_s$ is trained to recover the missing pixels based on information from the remaining pixels. The network's best guess would be the expected pixel value~\cite{krull2019noise2void}, and therefore the output image $\tilde{x}_{k}$ can be considered as the denoised version of $y_k$. 

The network $f_s$ is trained by minimizing the following loss function, which is the mean squared error (MSE) between the network output and the noisy image on the masked pixels.
\begin{equation}
    \mathcal{L}_s=\frac{1}{N+1}\sum_{k=0}^N || (1-b_k)\odot (\tilde{x}_k-y_k) ||_2^2
    \label{eqn:mask_mse}
\end{equation}

{\bf Spatial transformer network: } Let $\tilde{n}_k$ be the residual between the denoised image $\tilde{x}_k$ and the underlying clean image $x_k$, i.e., $\tilde{n}_k=\tilde{x}_k-x_k$, and thus $\tilde{x}_k=x_0\circ\phi_k^{-1}+\tilde{n}_k$. Suppose we can estimate $\phi_k$, the deformation field between frame $k$ and the target frame, and apply it to the denoised image $x_k$, then $\tilde{x}_k\circ\phi_k=x_0+\tilde{n}_k\circ\phi_k=x_0+\tilde{n}_k'$, i.e., $\tilde{x}_k\circ\phi_k$ is an image that is spatially matched with $x_0$ but is corrupted by noise $\tilde{n}'_k=\tilde{n}_k\circ\phi_k$. Note that $\{\tilde{x}_0, \tilde{x}_1\circ\phi_1, ..., \tilde{x}_N\circ\phi_N\}$ can be considered as a set of images that share the same spatial content but have different noise and can be used for multi-image denoising later.

We estimate the motion between the target frame and other frames using a STN module (Fig.~\ref{fig:method} b). A network $f_r$ is used to predict the deformation field given pairs of moving image $\tilde{x}_k$ and target image $\tilde{x}_0$, $\tilde{\phi}_k=f_r(\tilde{x}_k, \tilde{x}_0)$. Then $\tilde{x}_k$ is deformed with a grid sampler, $\tilde{x}_{k\rightarrow0}=\tilde{x}_k\circ\tilde{\phi}_k$. We adopt the architecture in~\cite{balakrishnan2019voxelmorph} for $f_r$ and optimize it to minimize the following loss function,
\begin{equation}
    \mathcal{L}_r=\frac{1}{N}\sum_{k=1}^N || \tilde{x}_{k\rightarrow0} -\tilde{x}_0 ||_2^2 + \lambda ||\nabla\tilde{\phi}_k||_2^2,
\end{equation}
where $\lambda$ is a weighting coefficient. The first term is the MSE between the warped image $\tilde{x}_k\circ\tilde{\phi}_k$ and the target image $\tilde{x}_0$, which is the similarity metric. The second term is the L2 norm of spatial gradient of $\tilde{\phi}_k$, serving as a regularization for the deformation field~\cite{balakrishnan2019voxelmorph}.

{\bf Multi-image denoising: } We now have two sets of images, the denoised and deformed images $\{\tilde{x}_0, \tilde{x}_{1\rightarrow0}, ..., \tilde{x}_{N\rightarrow0}\}$ and the original noisy images $\{y_0, y_1, ..., y_N\}$. In the final stage, we aggregate all these images to generate a refined estimation of the target clean image. We adopt the blind-spot method similar to the single-image denoising stage but concatenate all images as input and produce an estimation for the target frame, $\hat{x}_0 = f_m(\tilde{x}_0, \tilde{x}_{1\rightarrow0}, ..., \tilde{x}_{N\rightarrow0}, b\odot y_0, y_1, ..., y_N)$, 
where $\hat{x}_0$ is the final estimation and $f_m$ is the multi-image denoising network. Again, we use dropout to remove some pixels in $y_0$ to avoid learning a identity mapping. $f_m$ shares the same architecture with $f_s$ except for the number of input channels. Similar to Eq.~\ref{eqn:mask_mse}, a masked MSE loss is used to train network $f_m$, 
\begin{equation}
    \mathcal{L}_m=|| (1-b)\odot (\hat{x}_0-y_0) ||_2^2.
\end{equation}

\subsection{Training and inference}

As mentioned above, all the three modules in the proposed method is trained in unsupervised or self-supervised manners. Besides, we can train this pipeline end-to-end with gradient descent based methods. Another advantage of our framework is that it can be trained on a single noisy image (with several auxiliary observations) without large-scale external dataset. In summary, the total loss function for our proposed model is $\mathcal{L}=\lambda_s \mathcal{L}_s +\lambda_r \mathcal{L}_r + \mathcal{L}_m$, where $\lambda_s$ and $\lambda_r$ are weighting coefficients. During training, the networks are updated for $N_{train}$ iterations to learning a specific denoising model for the input images. At each iteration, input images are randomly rotated for data augmentation and the dropout layers sample different realizations of binary masks $\{b, b_0, ..., b_N\}$ so that the denoising networks can learn from different parts of the image.  

For inference, we run forward pass of the model for $N_{test}$ times with dropout layers enabled and average output $\hat{x}_0$'s to generate the final prediction.

\section{Experiments and Results}

\subsection{Datasets and experiments setup}

We use the following two datasets to evaluate our proposed denoising framework.

{\bf PINCAT}: The PINCAT phantom~\cite{sharif2007adaptive,lingala2011accelerated} simulates both cardiac perfusion dynamics and respiration with variability in breathing motion. The spatial matrix size of the phantom is $128\times128$, which corresponds to a resolution of 1.5 mm $\times$ 1.5 mm. The intensity of the phantom is normalized to $[0,1]$. The middle frame of the series is selected as the target frame.

{\bf ACDC}: The ACDC dataset~\cite{bernard2018deep} consists of short-axis cardiac cine MRI of 100 subjects. The in-plane resolution and gap between slices range from 0.7 mm$\times$0.7 mm to 1.92 mm$\times$1.92 mm and 5 mm to 10 mm respectively. For pre-processing, we linearly normalize the image intensity to $[0,1]$ and crop or pad them to a fixed size of $224\times224$. For each subject, we extract the sequence of the middle slice and use the end-systole (ES) and end-diastole (ED) phases as target frames. 

In all the experiments, we use another four noisy images from the same sequences as auxiliary observations (two adjecent frames before and after the target frame), i.e., $N=4$. We set $\lambda=0.1$, and $\lambda_s=\lambda_r=1$ in the loss function. For dropout layers, a dropout rate of 0.3 is used. We train the model using an Adam optimizer~\cite{kingma2017adam} with a learning rate of $1\times 10^{-4}$. $N_{train}$ is set to 2000 for PINCAT dataset and 4000 for ACDC dataset. $N_{test}$ is set to 100 for both datasets. The neural networks are implemented with PyTorch 1.5, and trained and evaluated on a Titan V GPU. For PINCAT dataset, we simulate Gaussian noise and Poisson noise at different noise levels. The standard deviation $\sigma$ of Gaussian noise is set to 15\%, 20\% and 25\%. The noisy observation $y$ under Poisson noise is generated by $y=z/P$, where $z\sim\mbox{Pois}(Px)$, $x$ is the truth intensity and $P$ is a parameter to control noise level. We set $P=40,20,$ and 10 in the experiments. For ACDC dataset, we simulate Gaussian noise and Rician noise~\cite{gudbjartsson1995rician}, with $\sigma=5\%, 10\%$, and 15\%. We compare our proposed method (D2S) with other state-of-the-art denoising methods, including deep image prior (DIP)~\cite{ulyanov2018deep}, Self2Self (S2S)~\cite{quan2020self2self}, BM3D~\cite{dabov2007image} and VBM4D~\cite{maggioni2012video}. For DIP and S2S, we use the same learning rate mentioned above and tune the number of iterations on our datasets. We adopt the peak signal to noise ratio (PSNR) and the structural similarity index measure (SSIM) as evaluation metrics. The reference PyTorch implementation for Deform2Self is available on GitHub\footnote{\url{https://github.com/daviddmc/Deform2Self}}

\subsection{Results}
{\bf Comparison with other approaches: } Table~\ref{tab:quant} shows the quantitative results on PINCAT and ACDC dataset. The proposed method achieves comparable or even better performance than other methods. Specifically, D2S has similar results to VBM4D in terms of Gaussian noise, and outperforms VBM4D for other noise models, especially for high noise levels. Besides, D2S outperforms S2S consistently, indicating that information from other time frames can largely boost performance of denoising models. 

\begin{figure}[t]
\centering
\includegraphics[width=\textwidth]{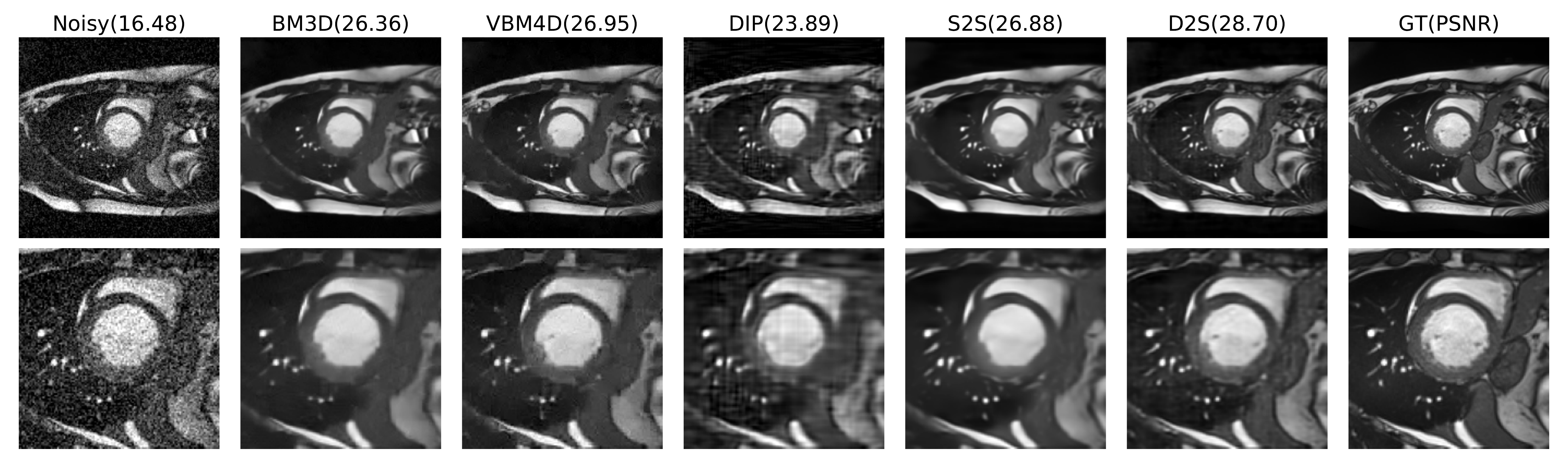}
\caption{Visualization results of different methods for ACDC dataset under Gaussian noise with $\sigma=15\%$.}
\label{fig:gaussian}
\end{figure}

\begin{figure}[t]
\centering
\includegraphics[width=\textwidth]{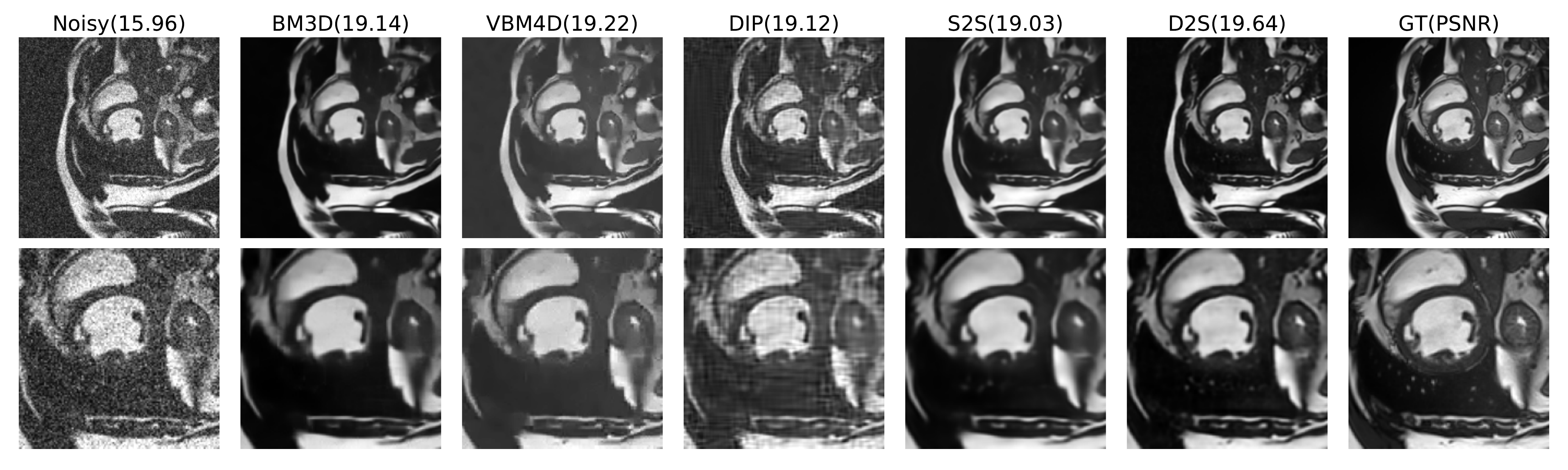}
\caption{Visualization results of different methods for ACDC dataset under Rician noise with $\sigma=15\%$.}
\label{fig:rician}
\end{figure}

Fig.~\ref{fig:gaussian} and~\ref{fig:rician} show example slices under Gaussian and Rician noise with $\sigma=15\%$ and the denoised results using different methods. The D2S method has not only better statistical but also better perceptual results compared with other methods. Single-image methods such as BM3D and S2S, only use single noisy image, and therefore have not enough information to recover details that are corrupted by noise, resulting in blurred estimation. The DIP method suffers from significant structural artifacts in high noise levels. Though retrieving some details from adjacent frames, VBM4D also brings subtle artifacts to the denoised images. D2S is able to recover more detail structures with higher image quality.

\begin{table}[h]
\centering
\begin{tabular}{c|cccccc|cccccc}
\toprule
\multicolumn{13}{c}{PINCAT}\\
\midrule
\multirow{3}{*}{Method} & \multicolumn{6}{c}{Gaussian} & \multicolumn{6}{c}{Poisson}\\
 & \multicolumn{2}{c}{$\sigma=15\%$} & \multicolumn{2}{c}{$\sigma=20\%$} & \multicolumn{2}{c}{$\sigma=25\%$} & \multicolumn{2}{c}{$P=40$} & \multicolumn{2}{c}{$P=20$} & \multicolumn{2}{c}{$P=10$} \\
\cline{2-13}
   & PSNR & SSIM & PSNR & SSIM & PSNR & SSIM & PSNR & SSIM & PSNR & SSIM & PSNR & SSIM\\
\midrule
Noisy     & 16.55 & 0.300 & 14.05 & 0.208 & 12.11 & 0.151 & 22.42 & 0.603 & 19.40 & 0.472 & 16.19 & 0.346 \\
BM3D      & 29.97 & 0.918 & 27.98 & 0.881 & 26.38 & 0.843 & 32.56 & 0.954 & 30.38 & 0.930 & 27.63 & 0.890 \\
VBM4D     & 31.36 & 0.936 & 29.65 & 0.913 & 28.28 & 0.886 & 32.35 & 0.953 & 29.92 & 0.930 & 27.65 & 0.899 \\
DIP       & 28.28 & 0.879 & 26.85 & 0.837 & 24.63 & 0.759 & 31.96 & 0.949 & 30.99 & 0.935 & 26.54 & 0.868 \\
S2S       & 30.27 & 0.928 & 28.04 & 0.900 & 27.68 & 0.883 & 33.05 & 0.962 & 31.25 & 0.951 & 30.55 & 0.939 \\
D2S       & \textcolor{red}{31.77} & \textcolor{red}{0.946} & \textcolor{red}{30.14} & \textcolor{red}{0.919} & \textcolor{red}{29.10} & \textcolor{red}{0.891} & \textcolor{red}{35.13} & \textcolor{red}{0.978} & \textcolor{red}{33.74} & \textcolor{red}{0.969} & \textcolor{red}{31.67} & \textcolor{red}{0.951} \\
\midrule
\multicolumn{13}{c}{ACDC}\\
\midrule
\multirow{3}{*}{Method} & \multicolumn{6}{c}{Gaussian} & \multicolumn{6}{c}{Rician}\\
 & \multicolumn{2}{c}{$\sigma=5\%$} & \multicolumn{2}{c}{$\sigma=10\%$} & \multicolumn{2}{c}{$\sigma=15\%$} & \multicolumn{2}{c}{$\sigma=5\%$} & \multicolumn{2}{c}{$\sigma=10\%$} & \multicolumn{2}{c}{$\sigma=15\%$} \\
\cline{2-13}
   & PSNR & SSIM & PSNR & SSIM & PSNR & SSIM & PSNR & SSIM & PSNR & SSIM & PSNR & SSIM\\
\midrule
Noisy     & 26.02 & 0.769 & 20.00 & 0.518 & 16.48 & 0.369 & 25.70 & 0.742 & 19.66 & 0.513 & 16.07 & 0.368  \\
BM3D      & 32.32 & 0.953 & 28.54 & 0.905 & 26.45 & 0.860 & 29.58 & 0.874 & 23.69 & 0.777 & 19.78 & 0.689 \\
VBM4D      & \textcolor{red}{32.54} & 0.957 & 28.96 & 0.911 & 26.88 & 0.863 & \textcolor{red}{29.79} & \textcolor{red}{0.879} & 23.94 & 0.791 & 19.93 & 0.707 \\
DIP       & 26.95 & 0.875 & 25.55 & 0.815 & 23.48 & 0.718 & 26.10 & 0.811 & 22.76 & 0.736 & 19.10 & 0.629 \\
S2S       & 30.41 & 0.942 & 28.45 & 0.912 & 26.90 & 0.880 & 28.28 & 0.861 & 23.51 & 0.784 & 19.73 & 0.709 \\
D2S       & 32.16 & \textcolor{red}{0.960} & \textcolor{red}{30.26} & \textcolor{red}{0.936} & \textcolor{red}{28.22} & \textcolor{red}{0.887} & 29.37 & \textcolor{red}{0.879} & \textcolor{red}{24.25} & \textcolor{red}{0.812} & \textcolor{red}{20.20} & \textcolor{red}{0.743} \\
\bottomrule
\end{tabular}
\caption{Quantitative results on PINCAT dataset for Gaussian and Poisson noise at different noise levels. The best results are indicated in red.}
\label{tab:quant}
\end{table}

{\bf Ablation study: } We perform ablation studies to evaluate different components in our framework. We evaluate models that ablate the single-image denoising module and STN module respectively. 

The ablated and full models are evaluated on the ACDC dataset with Gaussian and Rician noise ($\sigma=10\%$). To investigate how single-image denoising improve registration accuracy and how image registration help match image content for the following multi-image denoising, we compute evaluation metrics PSNR and SSIM on the region of interest (ROI) that involves cardiac motion. The ROI masks include left ventricle, myocardiam and right ventricle, which are annotated in the ACDC dataset~\cite{bernard2018deep}. 

Results of ablation studies (Table~\ref{tab:ablation}) indicate that the single-image denoising module improves the performance of the D2S model. It improves the accuracy of motion estimation in the registration network, and provides a good initialization for multi-image denoising. Besides, the registration module also makes contribution to the performance of our model. It registers images at different time frames to the same template so that the multi-channel input for the following denoising network is spatially matched, making the denoising task easier.

\begin{table}[!h]
\centering
\begin{tabular}{c|cccc}
\toprule
\multirow{2}{*}{Method} & \multicolumn{2}{c}{Gaussian ($\sigma=10\%$)} & \multicolumn{2}{c}{Rician ($\sigma=10\%$)}\\
 & ROI-PSNR & ROI-SSIM & ROI-PSNR & ROI-SSIM \\
\midrule
D2S     & 28.01 & 0.894 & 27.55 & 0.889\\
D2S w/o single-image denoising & 27.81 & 0.888 & 27.34 &  0.884\\
D2S w/o image registration & 27.68 &  0.887 & 27.24 &  0.883\\
\bottomrule
\end{tabular}
\caption{Quantitative results of ablation studies.}
\label{tab:ablation}
\end{table}

\section{Conclusions}

In this work, we proposed Deformed2Self, a self-supervised deep learning method for dynamic imaging denoising, which explores the similarity of image content at different time frames by estimating the motion during imaging and improve image quality with sequential single- and multi-image denoising networks. In addition, the proposed method only relies on the target noisy image with a small number of observations at other time frames and has no prerequisite on a large training dataset, making it more practical for applications with scarce data. Experiments on a variety of noise settings show that our method has comparable or even better performance than other state-of-the-art unsupervised or self-supervised denoising methods.

\bibliographystyle{splncs04}
\bibliography{ref.bib}




% LNCS Homepage, \url{http://www.springer.com/lncs}. Last accessed 4
% \end{thebibliography}
\end{document}